\documentstyle[12pt,aasms]{article}

\begin{document}

\title{\bf An Optical  Ultrahigh  Resolution Spectrograph for Use with  
Adaptive  Optics}

\author {J. Ge, B. Jacobsen, J.R.P. Angel,  N. Woolf, J. H. Black, M. Lloyd-Hart, P. Gray}
\affil{Center for Astronomical Adaptive Optics, Steward Observatory, University of Arizona, Tucson, AZ 85721}
\author{R. Q. Fugate}
\affil{Starfire Optical Range, U. S. Airforce Phillips Lab, Kirtland AFB, NM 87117}

\begin{abstract}
A prototype  ultrahigh resolution  
spectrograph has been built with  
  an adaptive optics
telescope. It  provides  
$250,000$ resolving power,  300 \AA\  wavelength coverage 
 and 0.8\% efficiency.  

\end{abstract}

\newpage
A prototype optical wavelength ultrahigh resolution echelle cross-dispersed 
spectrograph has been tested at the Starfire Optical Range (SOR) 1.5 m
telescope (Woolf et al. 1995; Ge et al. 1996). To our knowledge 
this is the  first high 
resolution spectrograph
to take advantage of the diffraction limited images produced by  an 
adaptive optics system.  
  Because of the  sharpened images produced by the adaptive optics at visible wavelength,  about r$_0$/D $
\sim 1/15$ in the seeing-dominant domain, the
 narrow slits necessary for high resolution can be used without a large loss of light. This is a great advantage when  compared with  conventional high resolution spectrographs (e.g.
Lambert et al. 1990; Diego et al. 1995). In addition, the smaller image widths 
inherent with the adaptive optics system allow the orders to be spaced 
closer together on the chip, allowing more orders to be observed simultaneously.

\smallskip

The main part of the spectrograph layout at the Coud\'{e}
room of the SOR  is shown in Figure 1. A $242\times 116$ mm$^2$ Milton Roy R2 echelle
grating with  23.2 gr mm$^{-1}$ and  a blaze angle of  63.5$^\circ$ 
 was used to 
provide the main dispersion. Theoretical resolution of this grating is 
 500,000 in the I band, 600,000 in the R band and 800,000
in the  V band.
Cross dispersion was provided by an  8$^\circ$ apex angle BK7 prism used in 
double pass configuration.
The grating was illuminated in  quasi-Littrow mode
with a small  ($\sim$0.25$^\circ$) out-of-plane tilt. 
Collimation was provided by a large off-axis parabola with a focal length 
of 6 m. 
 A large folding flat  was used to reduce the overall length of 
the spectrograph. A
Loral  back-illuminated   CCD with 15 $\mu$m pixels and 
antireflection (AR) coating
in a $2048\times 2048$ format was proposed. The pixel size was sufficient for 
Nyquist sampling. 
 The spectral format for the spectrograph is shown in Figure 2. 
The central box is the 
CCD's  physical size ($30\times 30$ mm$^2$), which can cover about 90 echelle
 orders, corresponding to  a wavelength range of  0.47 to 1 $\mu$m (or from 
the 163rd  to the 76th order), with a width of  $\sim$ 8 \AA\ per order. 
The 30 mm length of the CCD is such that
11 exposures are needed to obtain  a complete  coverage of all the  orders.

\smallskip

We have conducted two observation runs at the SOR 1.5 m telescope, one 
in June and  one in November of   1995.  Here we briefly
 report the results.

\smallskip

At the  SOR two different wavelength bands are available (but not 
simultaneously),  the $``$Blue" leg (0.47-0.7  $\mu$m) and  the $``$Red" leg 
(0.7-1.0 $\mu$m).
 Figure 3 and 4 show parts of spectra obtained for Vega for the Blue and Red legs
respectively. They were taken in  June 1995  with a   $2048\times 2048$ Kodak thermo-electrically 
cooled CCD with 9 $\mu$m  pixels. 
Figure 5 is a reduced
spectrum of a He-Ne laser obtained during the same run; 
the separate modes are clearly  resolved
indicating a resolving power of about 660,000 at 6328 \AA, which is very close
to the predicted resolution of 680,000. However, stellar observations
 from Vega and other bright stars give a much lower resolution (approximately 
250,000) due to the use of a larger slit (about 100 $\mu$m) (Figure 6).
Due to the use of a smaller Kodak CCD (because of electronics problems with the 
prepared Loral CCD), the wavelength coverages for the Blue and Red legs 
are approximately 300 \AA\ and 200 \AA, respectively. 

\smallskip

The spectrograph has 11 reflecting  and  10 transmitting  surfaces (including
a  relay system  before the slit). This leads  to large photon losses. 
Total efficiency for  the June run,  including
 sky and telescope transmission, spectrograph losses, 
and CCD quantum efficiency,
was only about 0.3\% near the  peak of the blaze. Improvements in coatings,
however, by November increased the total efficiency to  approximately
 0.8\% (Figure 7). 

\smallskip

Table 1 shows a comparison between different ultrahigh resolution 
spectrographs, and  includes the proposed spectrograph for the  SOR
3.5 m telescope.        
High resolution spectrographs mated with adaptive optics can make great 
 gains   in both throughput and wavelength coverage over 
 conventional spectrographs. 

\scriptsize
\begin{flushleft}
Table 1. Comparison between Different Ultrahigh Spectrographs in the World
\end{flushleft}
\begin{tabular}{lllll}  \hline
 \hline
 &&&&\\
Name  &McDonald (2.7 m)$^a$  &AAT (3.8 m) UHRF$^b$  &SOR 1.5 m & SOR 3.5 m (planned)\\
&&&&\\
Resolution&        5-6$\times 10^5$ & 9.9$\times 10^5$ &2.5$\times 10^5$$^f$
      & 7.7$\times 10^5$ \\
Total efficiency at V band  & 0.045\%  & 0.32\%          & 0.8\%         &
4\%$^g$\\
CCD &TI-2 $800\times 800$   & Thomson $1024\times 1024$& Kodak $2048\times 2048$  & Loral $4096\times 2048$ \\ 
Pixel Size ($\mu$m) &  15 & 19 & 9& 15\\
Readout Noise (e$^-$)& 15& 3.9 & 12.2 & 5\\
Pixel Number/Res. Element& 8 & 2.4 & 10.3 & 2  \\
Wavelength Coverage (\AA) & 1.2  & 2.5  & 200,270,500$^e$  & 1300 \\ 
Working Wavelength(\AA) & 3,300-8000$^c$ & 3,000-11,000 & 4,000-10,000 & 5,000-10,000\\
V Magnitude limit  (mag)$^d$& 4.4& 7.8&7.4&11.1\\ 
&&&&\\
\hline

\end{tabular}
\smallskip

\noindent $^a$ The information on McDonald Coud\'{e} Echelle is based on 
Hobbs \& Welty (1991) and Tull et al. (1995).

\noindent $^b$ The information on AAT UHRF is  based on Diego et al. (1995).

\noindent $^c$ From Cardelli (1995). 


\noindent $^d$ Based on the same readout noise, 5 e$^-$/pixel, same pixel 
size, 15 $\mu$m, 30 mins exposure and S/N = 30 per pixel.

\noindent $^e$ 200 \AA\ for Red Leg spectrum coverage (7,000-10,000 \AA) with
Kodak 2kx2k CCD with 9 $\mu$m pixels, 270 \AA\ for Blue Leg spectrum coverage (4,700-7,000 \AA) with
Kodak 2kx2k CCD. 500 \AA\ for Blue Leg spectrum coverage with Lesser's 
$2048\times 2048$  CCD  with 15 $\mu$m pixels. 

\noindent $^f$ Due to the slitwidth of a temporary slit ($\sim$ 100 $\mu$m), we
cannot narrow it down to the 40 $\mu$m width needed to reach more
than half million resolution.

\noindent $^g$ The efficiency has included the increase of the 3.5 m
 telescope/AO transmission, about 2.5 times   1.5 m telescope transmission
(Fugate 1995, private communication), and the increase of the  Loral
CCD QE, about 2 times the Kodak CCD QE. 

\begin{flushleft}
{\bf REFERENCES}
\end{flushleft}

\noindent Cardelli, J. A., 1995, in  Laboratory and Astronomical High Resolution Spectra, ASP Conf. Series, Vol. 81. 411.

\noindent Diego, F., et al. 1995, MNRAS, 272, 323

\noindent Ge, J., et al. 1996, in Preparation

\noindent Hobbs, L. M.,  \& Welty, D. E., 1991, ApJ, 368, 426

\noindent Kurucz, R. L., Furenlid, I., Brault, J., \& Testerman, L., 1984, Solar Flux Atlas from 296 to 1300 nm,

\noindent Lambert, D. L., Sheffer, Y., \& Crane, P., 1990, ApJ, 359, L19

\noindent  Tull, R. G., et al. 1995, PASP, 107, 251

\noindent Woolf, N., et al. 1995, Proc. SPIE Conference on Adaptive Optical Systems and Applications,  2534, Eds. R. K. Tyson \& R. Q. Fugate

\newpage

\normalsize
\centerline{\bf Figure Captions}

\noindent Figure 1 -- Optical layout of the SOR  cross-dispersed echelle spectrograph.

\noindent Figure 2 -- Echellogram for the SOR cross-dispersed echelle spectrograph. 
Wavelength increases from  left to right and from bottom to top. Several
order numbers and central wavelengths are marked. The central box
is the size for the Loral 2kx2k CCD with 15 $\mu$m square pixels. The order 
separations and central wavelengths are measured from the observation 
data obtained at SOR in June and November, 1995.

\noindent Figure 3 -- Vega spectra from SOR 1.5 m telescope/AO $``$Red" leg with SOR
Kodak 2kx2k CCD with 9 $\mu$m square pixels. Several 
orders and wavelengths are marked.

\noindent Figure 4 -- Vega spectra from SOR 1.5 m telescope/AO $``$Blue" leg with
the Kodak CCD. 

\noindent Figure 5 -- The spectral profile of the He-Ne laser at 632.8 nm. where
4 adjacent laser
modes separated by 0.00618 nm are clearly resolved.
The FWHM resolution is 3.9 pixels or 36 $\mu$m corresponding
to a resolving power of 660,000.

\noindent Figure 6 -- Part of the telluric O$_2$ absorption lines in the Vega spectrum 
obtained with the SOR echelle spectrograph. It is compared to the Kurucz et al. 
(1984) solar spectrum which was smoothed to resolution of 260,000 via 
convolution with Gaussian function. Two sets of  telluric 
lines show similar line profiles, indicating Vega spectral resolution of 
about 250,000.

\noindent Figure 7 -- Actually achieved total combined efficiency of the SOR echelle
spectrograph as a function of
 wavelength, which includes the sky transmission, telescope/AO transmission,
Strehl ratio of the image, slit loss and the spectrograph + Kodak CCD.  The 
quick fall-offs at wavelength longer than 8500 \AA\ and shorter than 5500 \AA\
are mainly caused by the fall-offs of the Kodak CCD QE at these wavelength ranges.

\newpage
\begin{figure}
\plotone{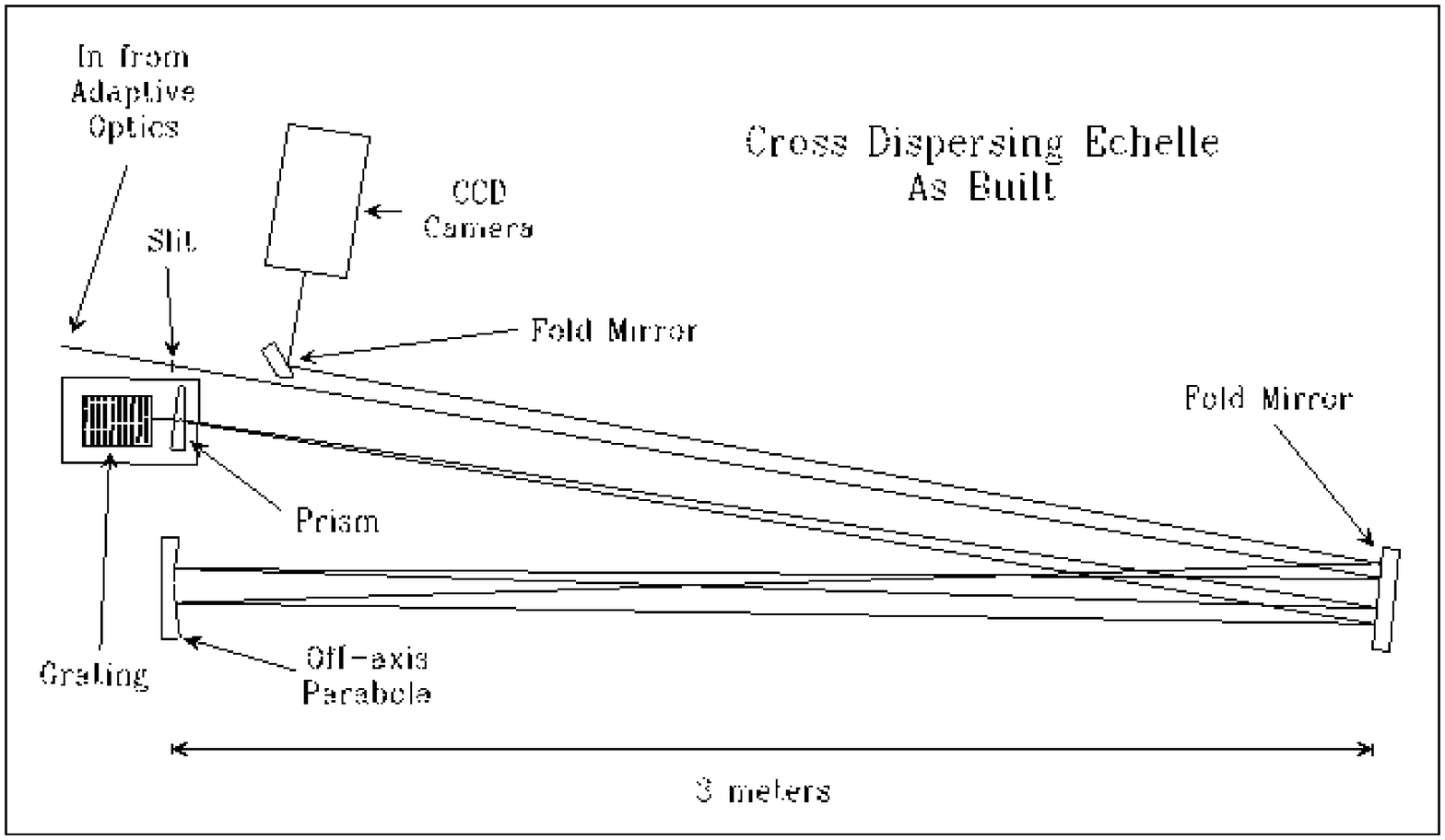}
\caption{}
\end{figure}

\newpage

\begin{figure}
\plotone{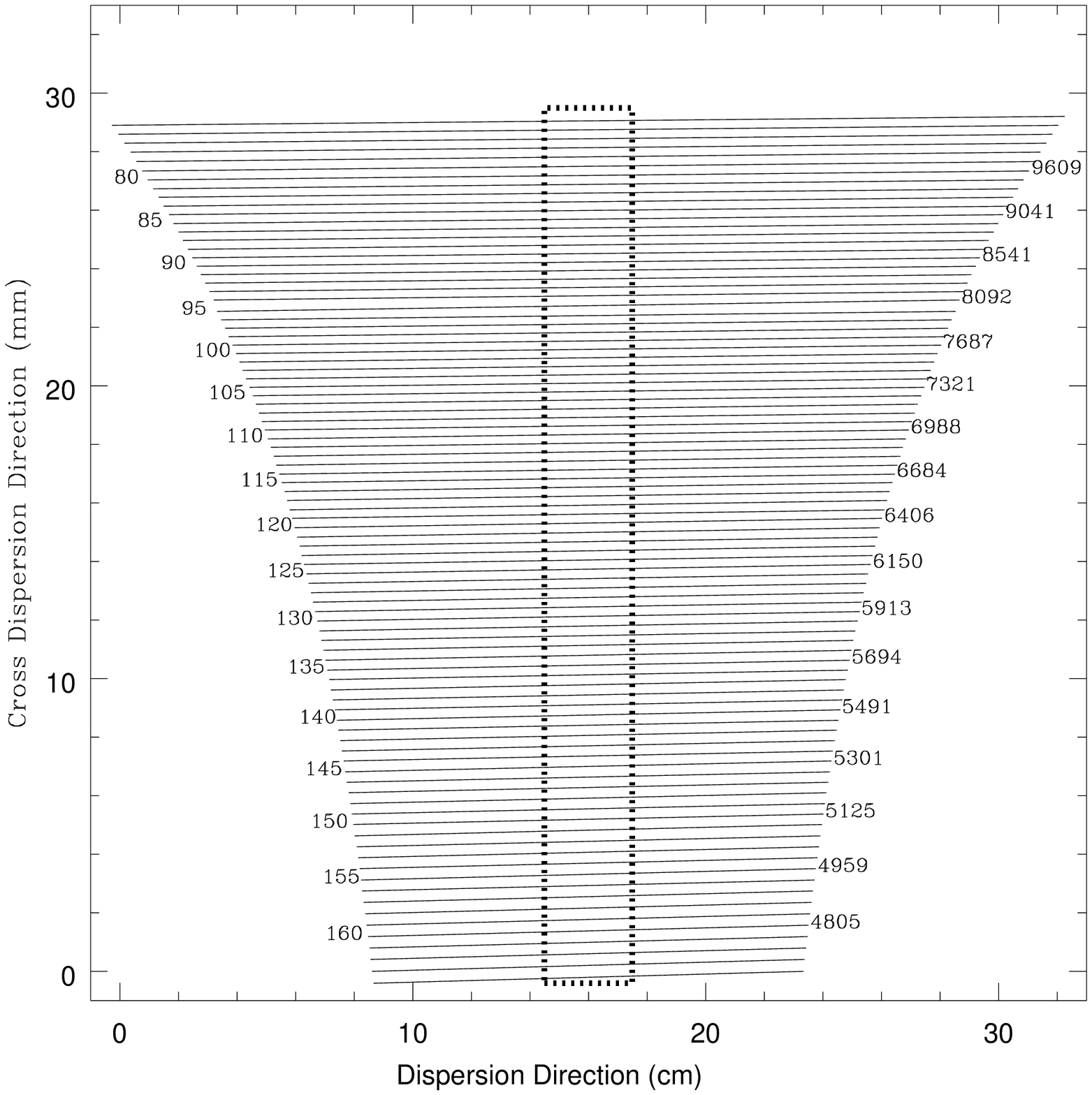}
\caption{}
\end{figure}

\newpage
\begin{figure}
\plotone{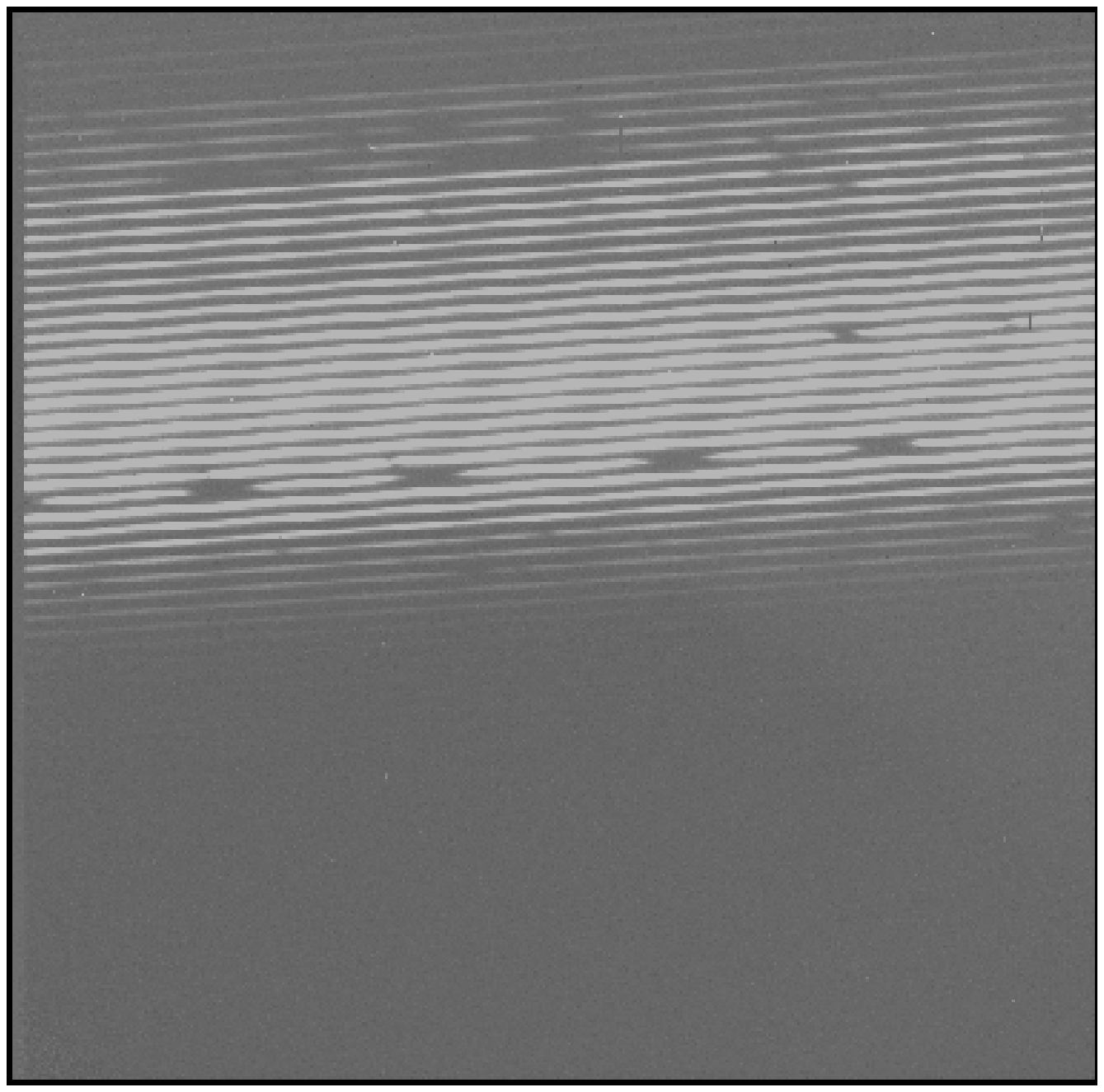}
\caption{}
\end{figure}

\newpage
\begin{figure}
\plotone{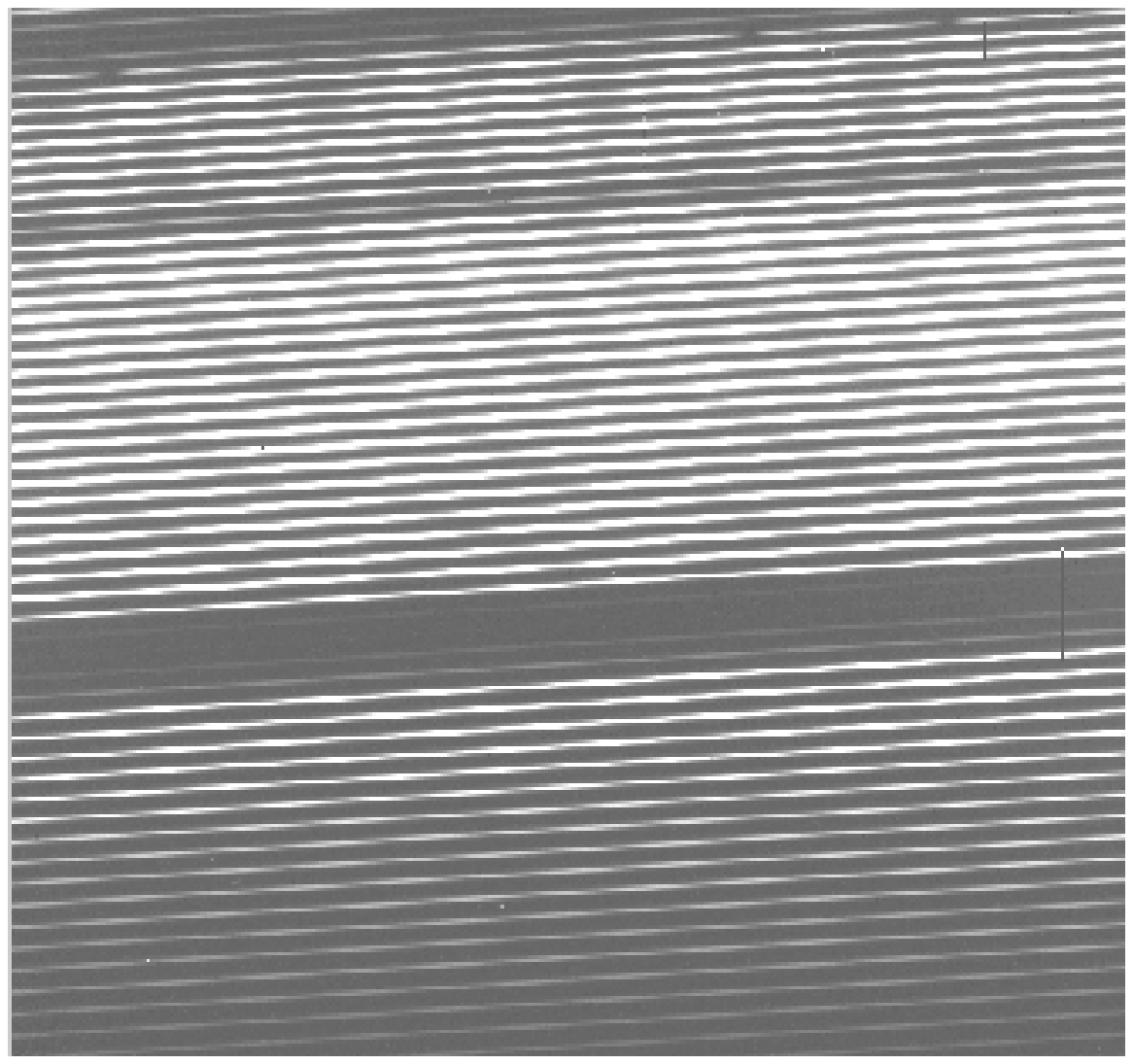}
\caption{}
\end{figure}

\newpage

\begin{figure}
\plotone{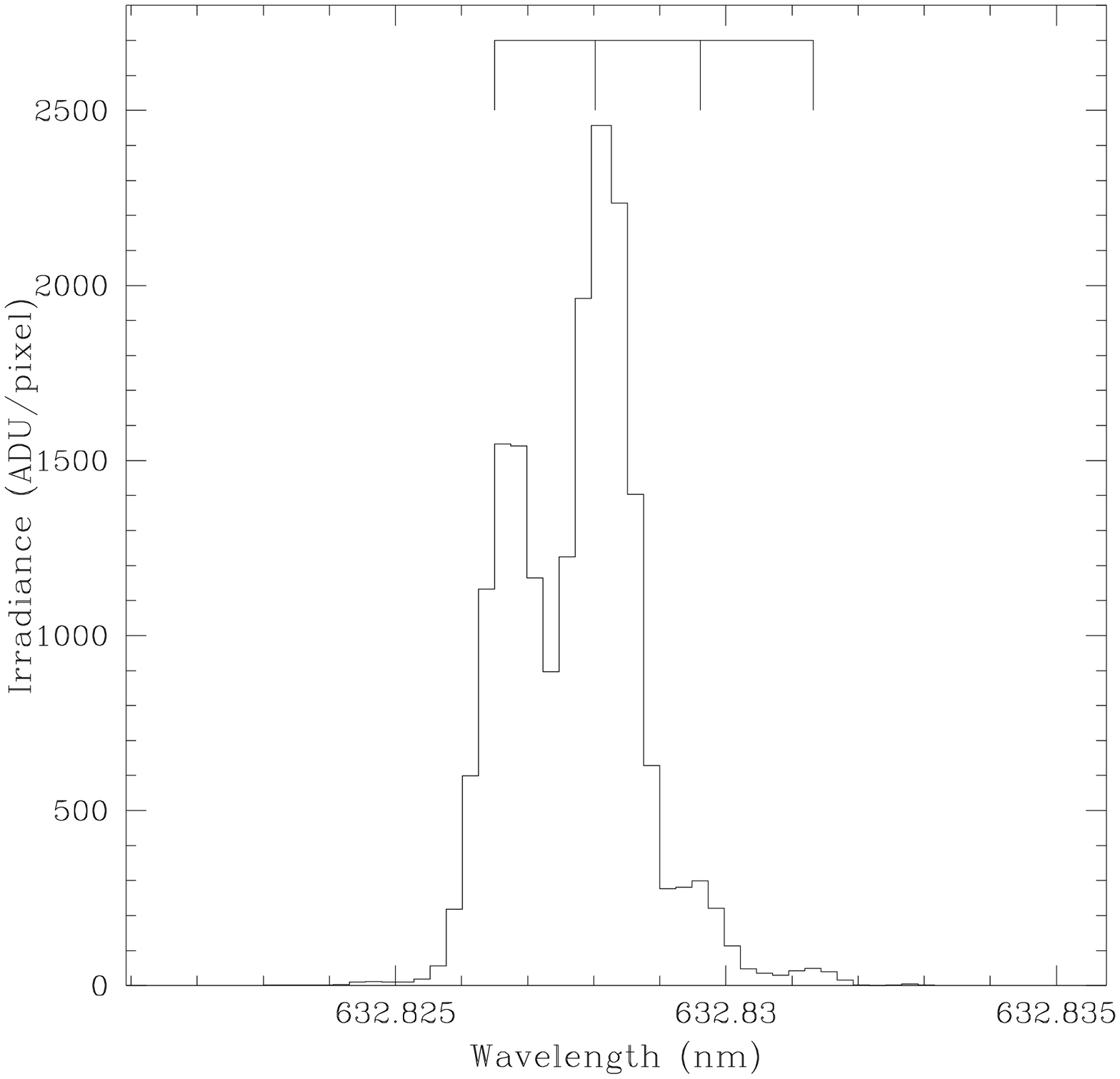}
\caption{}
\end{figure}

\newpage

\begin{figure}
\plotone{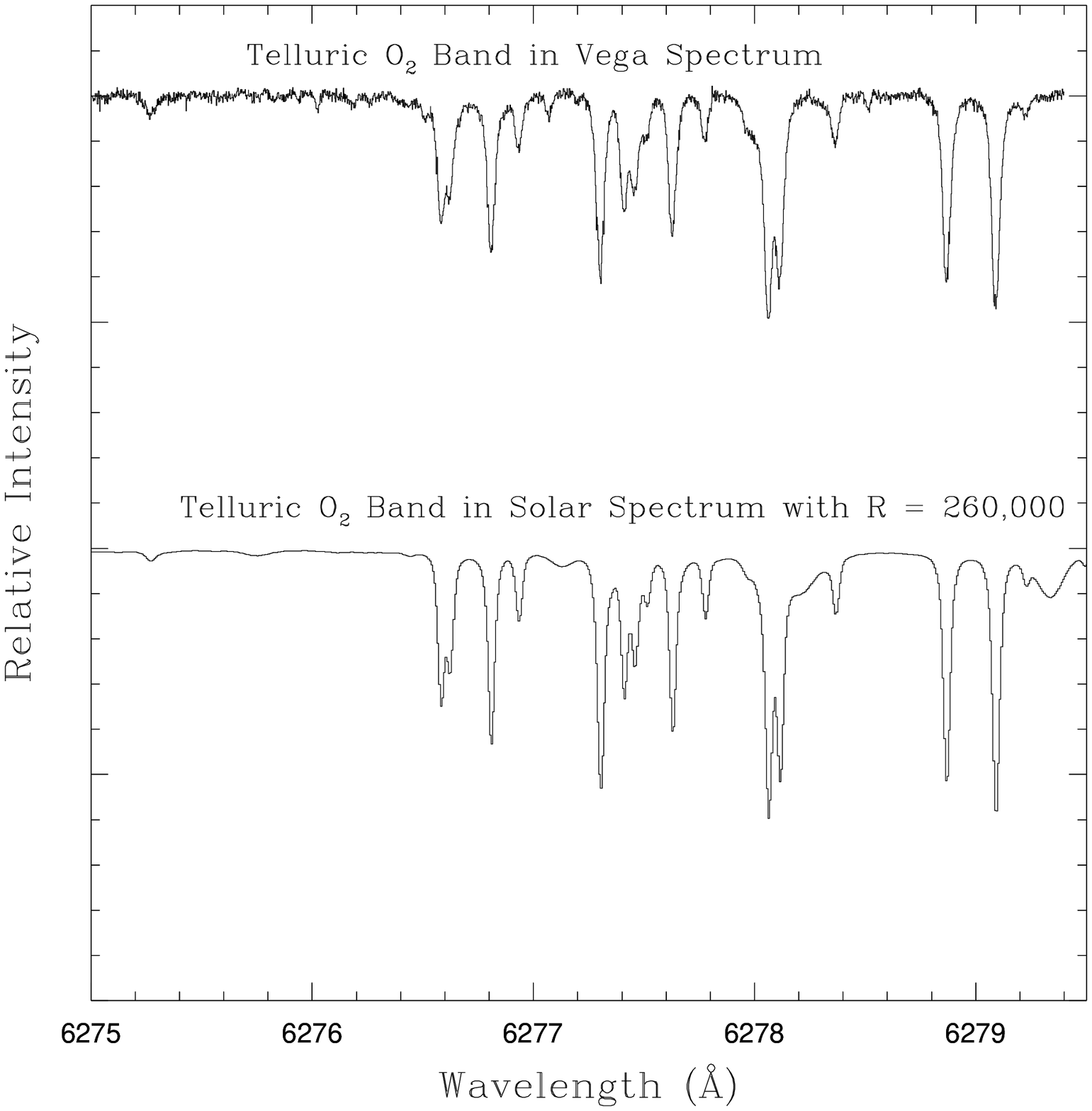}
\caption{}
\end{figure}

\newpage
\begin{figure}
\plotone{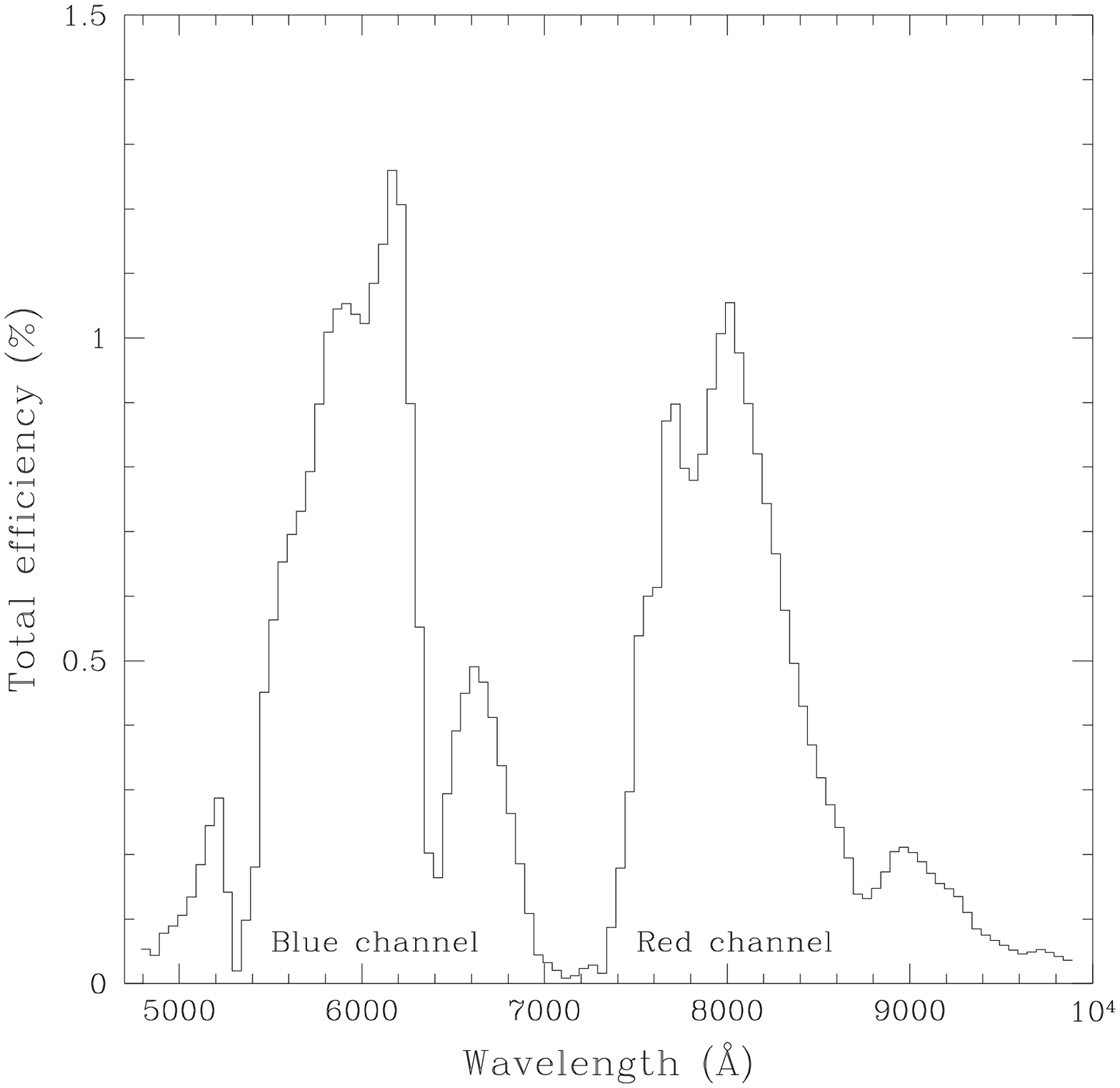}
\caption{}
\end{figure}

\end{document}